\documentclass{article}
\usepackage[preprint]{spconf}

\usepackage{amsmath}
\usepackage{graphicx}
\usepackage[labelformat=simple]{subcaption}

\usepackage[font=small,labelfont=bf]{caption}
\usepackage{multirow}
\usepackage{adjustbox}
\usepackage{amssymb}
\usepackage{bbm}
\usepackage{booktabs}
\usepackage{float}
\usepackage{tabu}
\usepackage{bm}
\usepackage{highlight}
\usepackage[noadjust]{cite}

\usepackage{amsthm}
\usepackage{color, colortbl}
\usepackage[subtle]{savetrees}

\usepackage{tikz}
\usetikzlibrary{tikzmark}

\usepackage{pifont}
\newcommand{\cmark}{\ding{51}}%
\newcommand{\xmark}{\ding{55}}%

\usepackage{tikzscale}
\usepackage{pgfplots}
\pgfplotsset{compat=1.9}
\pgfplotsset{
    myplotstyle/.style={
    legend style={draw=none, font=\small},
    legend cell align={left},
    legend pos=north east,
    ylabel style={align=center, font=\bfseries\boldmath},
    xlabel style={align=center, font=\bfseries\boldmath},
    x tick label style={font=\bfseries\boldmath},
    y tick label style={font=\bfseries\boldmath},
    scaled ticks=false,
    every axis plot/.append style={thick},
    },
}
\pgfplotsset{select coords between index/.style 2 args={
    x filter/.code={
        \ifnum\coordindex<#1\def\pgfmathresult{}\fi
        \ifnum\coordindex>#2\def\pgfmathresult{}\fi
    }
}}
\usetikzlibrary{matrix}
\usepgfplotslibrary{groupplots}

\definecolor{Gray}{gray}{0.9}
\definecolor{new_orange}{HTML}{F76811}
\definecolor{new_blue}{HTML}{115AF7}



\usepackage[T1]{fontenc}

\allowdisplaybreaks

\newcommand\dboxed[1]{\tikz [baseline=(boxed word.base)] \node (boxed word) [draw, rectangle, dashed, line cap=round] {#1};}

\newcommand{\bst}[1]{\textcolor[RGB]{27,158,119}{\textbf{#1}}}
\newcommand{\wst}[1]{\textcolor[RGB]{217,95,2}{\textbf{#1}}}

\title{Anchored Speech Recognition with Neural Transducers}
\name{Desh Raj\sthanks{Work done during internship at Meta AI.}$^1$, Junteng Jia$^2$, Jay Mahadeokar$^2$, Chunyang Wu$^2$, Niko Moritz$^2$, Xiaohui Zhang$^2$, Ozlem Kalinli$^2$}
\address{$^1$Center for Language and Speech Processing, Johns Hopkins University, USA, $^2$Meta AI, USA}
\usepackage[normalem]{ulem}
\newcommand{\stkout}[1]{\bgroup\markoverwith{\textcolor{#1}{\rule[0.5ex]{2pt}{0.6pt}}}\ULon}

\begin{document}

\setlength{\abovedisplayskip}{3pt}
\setlength{\belowdisplayskip}{3pt}

\copyrightnotice{\copyright\ IEEE 2023}

\maketitle
\begin{abstract}
Neural transducers have achieved human level performance on standard speech recognition benchmarks.
However, their performance significantly degrades in the presence of cross-talk, especially when the primary speaker has a low signal-to-noise ratio.
Anchored speech recognition refers to a class of methods that use information from an \textit{anchor segment} (e.g., wake-words) to recognize device-directed speech while ignoring interfering background speech.
In this paper, we investigate anchored speech recognition to make neural transducers robust to background speech.
We extract context information from the anchor segment with a \textit{tiny} auxiliary network, and use \textit{encoder biasing} and \textit{joiner gating} to guide the transducer towards the target speech.
Moreover, to improve the robustness of context embedding extraction, we propose auxiliary training objectives to disentangle lexical content from speaking style.
We evaluate our methods on synthetic LibriSpeech-based mixtures comprising several SNR and overlap conditions; they improve relative word error rates by 19.6\% over a strong baseline, when averaged over all conditions.
\end{abstract}

\vspace{-1em}
\section{Introduction}
\label{sec:intro}
\vspace{-0.5em}

Neural transducers (using RNNs or transformers)~\cite{Graves2012SequenceTW} have become the dominant modeling technique in end-to-end on-device speech recognition~\cite{He2019StreamingES, Wu2020StreamingTA, Li2020TowardsFA, Shangguan2019OptimizingSR}, since they allow streaming transcription similar to CTC models~\cite{Graves2006ConnectionistTC, Zhang2021BenchmarkingLC, Li2019ImprovingRT}, while still retaining conditional dependence, like attention-based encoder-decoders (AEDs)~\cite{Chan2016ListenAA,Kim2017JointCB}. Although they have shown state-of-the-art performance on several benchmarks~\cite{Liu2021ImprovingRT}, they still suffer from degradation caused by interference due to background speech and noise~\cite{Andrusenko2020TowardsAC,Shi2021ImprovingRT}. Recent studies have used context audio for \textit{implicit} speaker and environment adaptation of transducer models~\cite{Schwarz2021ImprovingRA}.

In this paper, we focus on the problem of suppressing background speech using \textit{explicit} auxiliary information (often referred to as target speech extraction/recognition in literature~\cite{Sato2021MultimodalAF,Moriya2022StreamingTA}). Such auxiliary information is usually provided in the form of speaker embeddings (e.g., d-vectors in VoiceFilter-Lite~\cite{Wang2020VoiceFilterLiteST,Rikhye2021MultiUserVV}) or enrollment utterances (e.g., SpeakerBeam~\cite{Delcroix2018SingleCT, molkov2021AuxiliaryLF}). However, these strategies require the target speaker to be \textit{enrolled} with the device, which may not always be feasible or desirable from a privacy perspective. In contrast, \textit{anchored speech recognition} refers to a class of methods that use information from an \textit{anchor segment} (such as a wake-word) to recognize device-directed speech. By relying only on the anchor segment and extracting the auxiliary information on-the-fly, these models bypass the need for a speaker enrollment stage. The idea was first proposed in the context of hybrid ASR systems \cite{King2017RobustSR} and later extended to AED models using a speaker encoder network to extract auxiliary information from the anchor segment~\cite{Wang2019EndtoendAS}. 

We investigate anchored speech recognition to improve the performance of transducers in the presence of background speech.
In particular, we add a \textit{tiny} auxiliary network to extract context information from the anchor segment, and use it to bias the transducer towards the primary speaker.
In order to disentangle speaking style from lexical content in the context embedding, we explore several auxiliary training objectives.
We conduct controlled evaluations on LibriSpeech mixtures, where our models show relative word error rate (WER) improvements of 19.6\%, on average, compared to an Emformer baseline trained with background augmentation.
%


\vspace{-1em}
\section{Anchored speech recognition}
\vspace{-0.5em}

\begin{figure}
    \centering
    \includegraphics[width=0.8\linewidth]{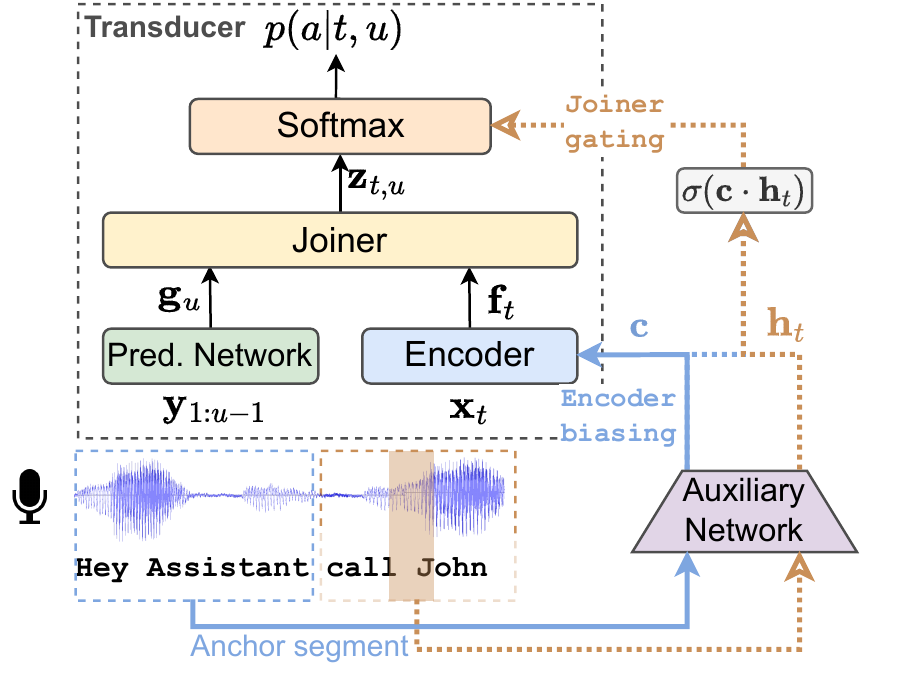}
    \vspace{-0.5em}
    \caption{
    Overview of transducer-based anchored speech recognition.
    }
    \label{fig:ts_asr}
\vspace{-1.6em}
\end{figure}

\subsection{Preliminary: ASR with neural transducers}
\vspace{-0.5em}

%
Given an utterance $\mathbf{x} = (\mathbf{x}_1,\ldots,\mathbf{x}_T)$, where $\mathbf{x}_t \in \mathbb{R}^d$ denotes audio features, transducers model the conditional probability of the output sequence $\mathbf{y} = (y_1,\ldots,y_U)$, where $y_u \in \mathcal{Y}$ denotes output units such as graphemes or word-pieces.
This is achieved by marginalizing over the set of all alignments $\mathbf{a} \in \bar{\mathcal{Y}}^{\ast}$, where $\bar{\mathcal{Y}} = \mathcal{Y}\cup \{\phi\}$ and $\phi$ is called the blank label.
Formally,
\begin{equation}
P(\mathbf{y}|\mathbf{x}) = \sum_{\mathbf{a}\in \mathcal{B}^{-1}(\mathbf{y})} P(\mathbf{a}|\mathbf{x}),
\label{eq:rnnt}
\end{equation}
where $\mathcal{B}$ is a deterministic mapping from an alignment $\mathbf{a}$ to an output sequence $\mathbf{y}$.
As shown in the \dboxed{box} in Fig.~\ref{fig:ts_asr}, transducers parametrize $P(\mathbf{a}|\mathbf{x})$ with an encoder, a prediction network, and a joiner.
The encoder maps $\mathbf{x}$ into hidden representations $\mathbf{f}_{1:T}$, while the prediction network maps $\mathbf{y}$ into $\mathbf{g}_{1:U}$.
The joiner combines the outputs from the encoder and the prediction network to compute logits $\mathbf{z}_{t,u}$ which are fed to a softmax function to produce a posterior distribution over $\bar{\mathcal{Y}}$.

\vspace{-0.8em}
\subsection{Biasing with the anchor segment}
\vspace{-0.5em}

Under the \textit{anchored speech recognition} framework, we assume that the model is deployed on a device where each utterance is preceded by an anchor segment (e.g., ``Hey Assistant'') that is relatively clean.
We extract context information from this anchor segment and use it to guide the transcription of the main utterance.
Assuming that the anchor segment is the first $T_w$ frames of the utterance $\mathbf{x}$, we modify equation~(\ref{eq:rnnt}) as
\begin{equation}
P(\mathbf{y}|\mathbf{x}) = \sum_{\mathbf{a}\in \mathcal{B}^{-1}(\mathbf{y})} P(\mathbf{a}|\mathbf{x},T_w).
\end{equation}

Next, we describe our proposed methods for biasing the transducer using a jointly trained auxiliary network to provide context information from the anchor segment.


\vspace{-0.8em}
\section{How to bias the transducer?}
\vspace{-0.5em}

Interference due to the presence of background speech may result in two categories of errors: (i) substitution/deletion errors caused by overlapping speech, and (ii) insertion errors from transcribing the undesired background speech. We propose two complementary methods for tackling these issues.

\vspace{-0.8em}
\subsection{Encoder biasing}
\vspace{-0.5em}

We extract a context embedding $\mathbf{c} = \textsf{Aux}(\mathbf{x}_{1:T_w}) \in \mathbb{R}^D$ from the anchor segment and use it to bias the encoder of the transducer, i.e., $\mathbf{f}_{1:T} = \textsf{Enc}(\mathbf{x}, \mathbf{c})$.
%
%
By adding the anchor segment embedding, we provide the encoder extra information to extract primary speech (or suppress overlapping background speech) in the hidden representations $\mathbf{f}_t$. This context information encourages the model to focus on speech with acoustics similar to the anchor segment, potentially reducing both type (i) and (ii) errors.

In our experiments, we use encoders which consist of time-stacked log Mel filterbanks (with 4x subsampling) fed into an Emformer~\cite{Shi2021EmformerEM}. By slight abuse of notation, let $\mathbf{x}_{1:T}$ refer to this down-sampled feature sequence (instead of the original input sequence). We found that a strategy of concatenating $\mathbf{c}$ to the time-stacked features followed by a non-linear projection works well\footnote{We tried other biasing techniques such as dynamic layer normalization~\cite{Kim2017DynamicLN} and context-aware query in the Emformer, but they did not perform well.}. Formally, this means that the input to the Emformer (originally $\mathbf{x}_t$) is now given as $\mathbf{x}^{\prime}_t = \textsf{ReLU}([\mathbf{x}_t^\mathsf{T}, \mathbf{c}^\mathsf{T}]^\mathsf{T}\mathbf{W}_{\mathrm{proj}})$, where $\mathbf{W}_{\mathrm{proj}} \in \mathbb{R}^{(d+D)\times d}$ is a projection matrix.

\vspace{-0.5em}
\subsection{Joiner gating}
\vspace{-0.5em}

In principle, encoder biasing should be sufficient for suppressing background speech in regions where the main speaker is inactive. However, we found that it still allows some cross-talk to leak, particularly in low SNR conditions. To alleviate the potential insertion error, we propose joiner gating which explicitly boosts the prediction of silence in non-target regions\footnote{The term ``gating'' here is not used in the standard sense of multiplicative gating, although that can be considered potential future work.}. Let $\mathbf{x}_{\Delta t}$ denote a small sub-segment of the input $\mathbf{x}$ at time $t$ with some left and right context. We obtain $\mathbf{h}_t=\textsf{Aux}(\mathbf{x}_{\Delta t})$ and compute a per-frame similarity bias $\mathbf{b}_t = \varphi(\mathbf{c},\mathbf{h}_t)$, where $\varphi: \mathbb{R}^D\times\mathbb{R}^D \rightarrow [0,1)$ is a function that increases monotonically with increasingly similar inputs. This bias is added to the non-blank label logits of the joiner output, and its complement is added to the blank logit. Formally, the logits after gating are given as
\begin{equation}
    \hat{\mathbf{z}}_{t,u} = \begin{cases}
        \mathbf{z}_{t,u} + (1 - \mathbf{b}_t)\quad \mathrm{if}~~\bar{y}_u = \phi, \\
        \mathbf{z}_{t,u} + \mathbf{b}_t \quad \quad \quad ~~ \mathrm{otherwise}.
    \end{cases}
\end{equation}
Here, we used $\varphi(\mathbf{c},\mathbf{h}_t) = \sigma\left(\frac{\mathbf{c}\cdot \mathbf{h}_t}{\lVert\mathbf{c}\rVert \lVert\mathbf{h}_t\rVert}\right)$, where $\sigma(\cdot)$ is the sigmoid function. Since we use function merging~\cite{Li2019ImprovingRT} to compute RNN-T loss directly from the logits, and since layer-norm is applied on the encoder and predictor outputs, it is both tractable and reasonable to add the bias to the logits instead of to the softmax output.

\begin{figure}
\begin{subfigure}{0.49\linewidth}
\centering
\includegraphics[width=\linewidth]{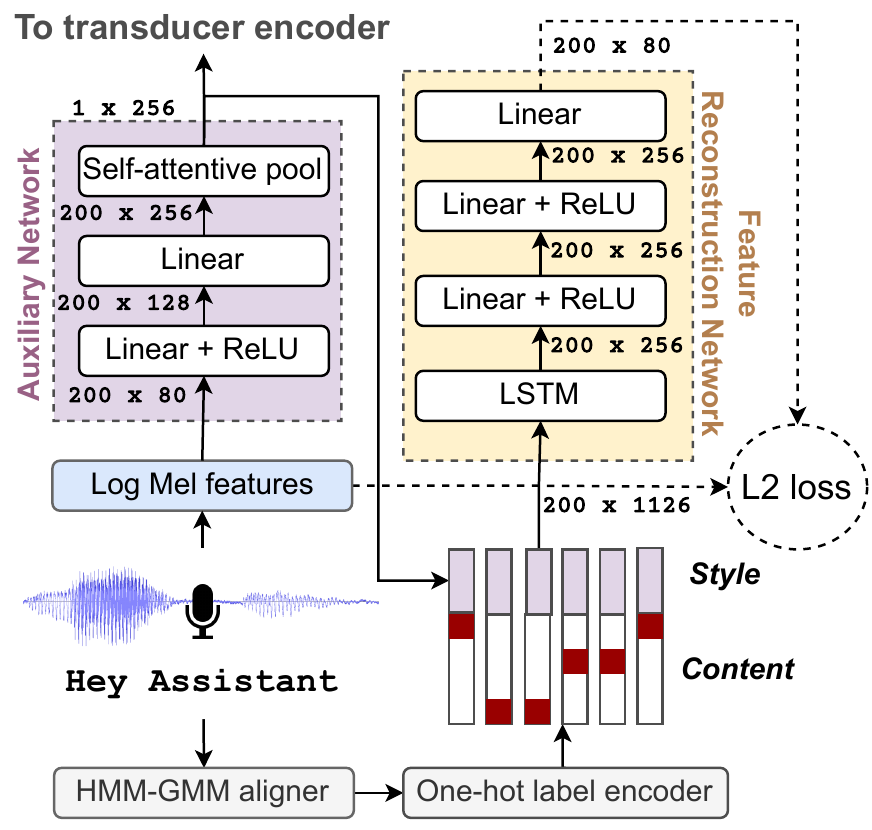}
\caption{}
\label{fig:fr}
\end{subfigure}
\begin{subfigure}{0.49\linewidth}
\centering
\includegraphics[width=\linewidth]{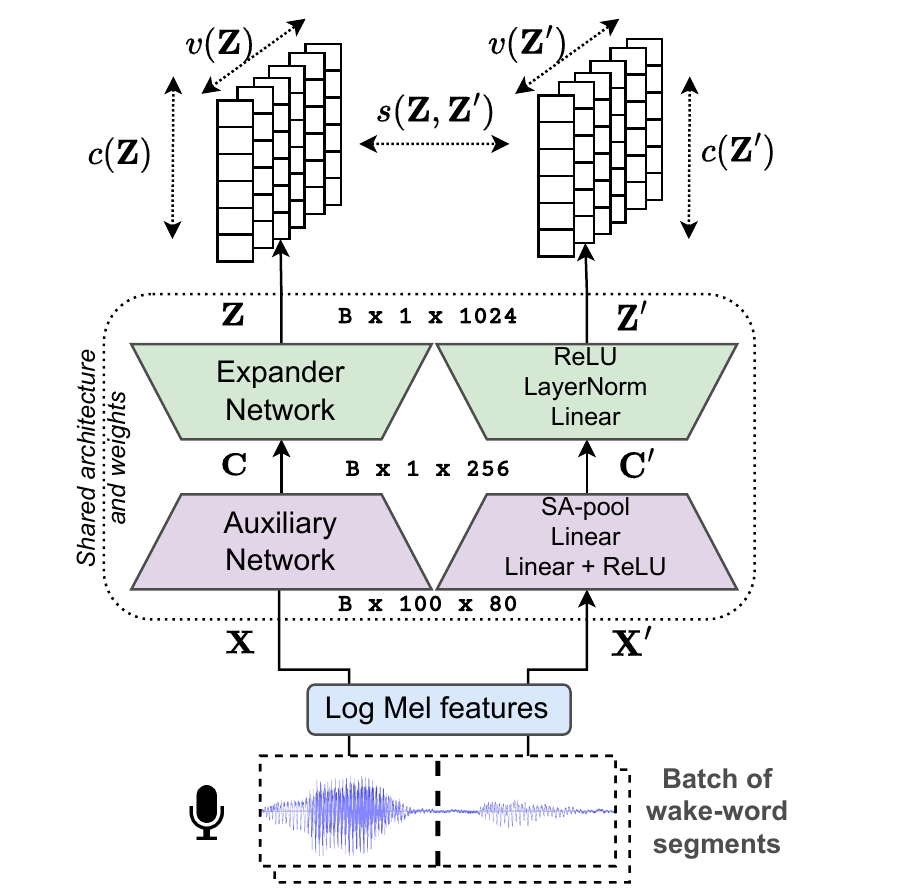}
\caption{}
\label{fig:vic_regularization}
\end{subfigure}
\vspace{-1em}
\caption{(a) Feature reconstruction and (b) VIC regularization as auxiliary objectives. FR uses frame-aligned chenone labels obtained from an HMM-based aligner, whereas VIC does not require such labels.}
\label{fig:auxiliary}
\vspace{-1.6em}
\end{figure}

\vspace{-0.8em}
\section{Disentangling style from content}
\vspace{-0.5em}

Training the model to recognize only the anchored speech should incentivize the auxiliary network to extract only speaking style characteristics from the anchor segment. 
%
%
%
Nevertheless, we employ auxiliary objective functions (shown in Fig.~\ref{fig:auxiliary}) to further suppress lexical content information in the context embedding $\mathbf{c}$.

\vspace{-0.8em}
\subsection{Feature reconstruction}
\vspace{-0.5em}

Let $\mathbf{s} \in \mathbb{Z}^{T_w}$ denote the frame-level estimated chenone (or senone) sequence corresponding to the anchor segment $\mathbf{x}_{1:T_w}$. The context embedding $\mathbf{c}$ and the chenone labels $\mathbf{s}$ are used to reconstruct the anchor segment using a feature reconstruction network similar to~\cite{Stafylakis2019SelfsupervisedSE} (shown in the yellow box in Fig.~\ref{fig:fr}), and the network is jointly trained using mean squared error:
\begin{equation}
\mathcal{L}_{\mathrm{FR}} = \textsf{MSE}(\mathbf{x}, \hat{\mathbf{x}}), \quad \hat{\mathbf{x}} = g_{\mathrm{FR}}(\mathbf{s},\mathbf{c};\Theta_{\mathrm{FR}}),
\end{equation}
where $g_{\mathrm{FR}}$ is the feature reconstruction (FR) network.


\vspace{-0.8em}
\subsection{VIC regularization}
\vspace{-0.5em}

Alternatively, we can penalize the auxiliary network for encoding lexical content by ensuring that the context embeddings generated from different parts of the anchor segment are similar.
This penalty is enforced through a non-contrastive self-supervised learning technique called VIC (variance-invariance-covariance) regularization, originally proposed in~\cite{Bardes2021VICRegVR} and shown in Fig.~\ref{fig:vic_regularization}.
Given a batch of anchor segments, we split it into two parts ($\mathbf{X}$, $\mathbf{X}^{\prime}$) by length and obtain the corresponding context embeddings ($\mathbf{C}$, $\mathbf{C}^{\prime} \in \mathbb{R}^{N\times D}$).
An \textit{expander network} projects them into higher dimensional representations ($\mathbf{Z}$, $\mathbf{Z}^{\prime} \in \mathbb{R}^{N\times \mathfrak{D}}$).
The VIC regularization objective is then given as
\begin{equation}
\mathcal{L}_{\mathrm{VIC}} = \gamma (\underbrace{v(\mathbf{Z})+v(\mathbf{Z}^{\prime})}_{\text{variance}}) + \mu \underbrace{s(\mathbf{Z},\mathbf{Z}^{\prime})}_{\text{invariance}}
+ \nu (\underbrace{c(\mathbf{Z})+c(\mathbf{Z}^{\prime})}_{\text{covariance}})
\end{equation}
where the terms are defined as
\begin{align*}
v(\mathbf{Z}) &= \frac{1}{\mathfrak{D}}\sum_{j=1}^{\mathfrak{D}} \max (0, 1-\sqrt{\mathrm{Var}(\mathbf{z}^j)}, \\
s(\mathbf{Z},\mathbf{Z}^{\prime}) = &\textsf{MSE}(\mathbf{Z},\mathbf{Z}^{\prime}),~~\mathrm{and}~~ \textbf{} c(\mathbf{Z}) = \frac{1}{\mathfrak{D}}\sum_{i\neq j}[C(\mathbf{Z})]_{i,j}^2,
\end{align*}
and $C(\mathbf{Z})$ is the covariance matrix for $\mathbf{Z}$. The invariance term pulls the representations for the two halves closer, while the variance term prevents them from collapsing to a constant. The covariance term ensures that the dimensions are decorrelated so as to increase information content. The overall training objective is then given as
\begin{equation}
\mathcal{L} = \mathcal{L}_{\mathrm{RNNT}} + \mathcal{L}_{\mathrm{VIC}}.
\end{equation}
Unlike the feature reconstruction loss, this method does not require frame-level chenone labels during training.

\vspace{-1em}
\section{Experimental Setup}
\vspace{-0.5em}

\subsection{Data}
\vspace{-0.5em}
\label{sec:data}

\begin{figure}[t]
    \centering
    \includegraphics[width=\linewidth]{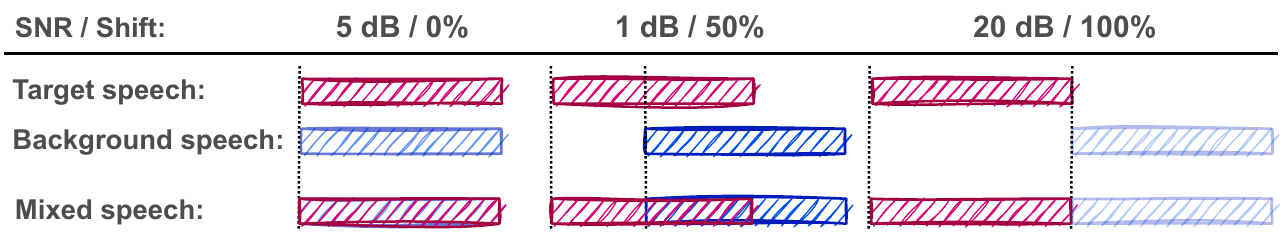}
    \caption{Illustration of mixture simulation for all shift conditions (and arbitrary SNR conditions).}
    \label{fig:data_mixing}
    \vspace{-1.5em}
\end{figure}

To enable a granular evaluation, we created synthetic mixtures (\texttt{dev} and \texttt{test}) using LibriSpeech~\cite{Panayotov2015LibrispeechAA} utterances (\texttt{clean} + \texttt{other}) as follows. For each (main) utterance, we randomly sampled another (background) utterance and cropped/tiled it to the length of the main utterance. The two were then mixed with a specific SNR (chosen from \{1, 5, 10, 20, 50\} dB) and shifts (chosen from \{0, 50, 100\}\%), as shown in Fig.~\ref{fig:data_mixing}.

This simulation resulted in 15 evaluation sets (5 SNR conditions $\times$ 3 shift conditions), each containing 5564 and 5556 utterances for the \texttt{dev} and \texttt{test} sets, respectively. Although we are unable to show results using real device-directed speech due to internal privacy regulations, the synthetic mixtures are still meaningfully similar. For training, we randomly mixed utterances from the 960h train set on-the-fly with 50\% probability, a fixed SNR of 10 dB, and a randomly sampled shift between 0\% and 100\%. Since LibriSpeech does not contain an explicit anchor segment, we fixed the first 2 seconds of the utterance as the anchor. The 0\% shift condition illustrates the scenario when the anchor segment is also noisy. During training, the wake-word segment was extracted from the clean or mixed utterances with 80\% and 20\% probability, respectively.

\vspace{-0.8em}
\subsection{Baselines}
\vspace{-0.5em}

We compared our models against 20-layer (base) and 24-layer (large) Emformer-transducer~\cite{Shi2021EmformerEM} trained with background speech augmentation as described above. Additionally, since our model uses the anchor segment to suppress background speech, we implemented simple anchor mean baselines: (i) subtraction (AMS), based on \cite{King2017RobustSR}, where the feature-level mean of the anchor segment is subtracted from the utterance, and (ii) concatenation (AMC), where the mean is appended to the input features, followed by an affine transformation.\footnote{AMS can be considered a special case of AMC when the transformation matrix is $[I_d,-I_d]^T$, where $I_d$ is the identity matrix.}

\begin{table*}[t]
\caption{
Comparison of proposed models with baselines on LibriSpeech-based \texttt{test} set, for different SNR and shift conditions, in terms of WER~(\%). Our main baseline is the Emformer-base model trained with background speech augmentation, denoted by ($\boldsymbol{\Lambda}$). The last column shows relative WER reduction (WERR) compared to $\boldsymbol{\Lambda}$, averaged across all conditions. WERs in \bst{green} are 5+\% better, and those in \wst{red} are 5+\% worse, than $\boldsymbol{\Lambda}$, in relative terms. $^\dagger$Extra parameters are only required during training.
}
\label{tab:results}
\vspace{-0.5em}
\adjustbox{max width=\textwidth}{%
\begin{tabular}{@{}lc@{\hskip 20pt}ccccc@{\hskip 20pt}ccccc@{\hskip 20pt}cccccr@{}}
\toprule
\multirow{2}{*}{\textbf{Model}} & \multirow{2}{*}{\textbf{Size(M)}} & \multicolumn{5}{c@{\hskip 20pt}}{\textbf{Shift = 0\%}} & \multicolumn{5}{c@{\hskip 20pt}}{\textbf{Shift = 50\%}} & \multicolumn{5}{c}{\textbf{Shift = 100\%}} & \multirow{2}{*}{\textbf{WERR}} \\
\cmidrule(r{20pt}){3-7} \cmidrule(r{20pt}){8-12} \cmidrule(l{-1pt}){13-17}
& & \textbf{1} & \textbf{5} & \textbf{10} & \textbf{20} & \textbf{50} & \textbf{1} & \textbf{5} & \textbf{10} & \textbf{20} & \textbf{50} & \textbf{1} & \textbf{5} & \textbf{10} & \textbf{20} & \textbf{50} \\
\midrule
Emformer-base (20L) ($\boldsymbol{\Lambda}$) & 76.7 & 50.28 & 18.42 & 10.12 & 7.82 & 7.27 & 28.49 & 11.46 & 7.88 & 7.13 & 7.02 & 65.71 & 38.74 & 14.85 & 7.32 & 7.03 & 0.0\% \\
Emformer-large (24L)  & 89.4 & 49.36 & 17.03 & 9.65 & 7.56 & 7.02 & 22.14 & 9.59 & 7.46 & 6.98 & 6.80 & 55.09 & 29.35 & 11.24 & 7.00 & 7.05 & 9.3\% \\
\midrule
Anchor mean sub.(AMS)~\cite{King2017RobustSR} & 76.7 & \wst{59.16} & \wst{27.48} & \wst{15.00} & 8.21 & \wst{7.93} & \bst{10.02} & \bst{8.30} & \bst{7.46} & 7.15 & \wst{7.71} & \bst{11.24} & \bst{8.92} & \bst{7.90} & \wst{7.69} & \wst{7.69} & 10.1\% \\
Anchor mean concat. (AMC) & 76.8 & 50.45 & \bst{16.95} & \bst{9.55} & 7.64 & 7.21 & \bst{17.09} & \bst{9.06} & 7.51 & 7.06 & 7.00 & \bst{42.74} & \bst{19.83} & \bst{8.03} & 6.99 & 6.99 & 14.5\% \\
\midrule
Encoder biasing & 77.2 & 52.65 & 19.23 & 9.82 & 7.59 & 7.12 & \bst{13.20} & \bst{8.67} & 7.57 & 7.12 & 6.92 & \bst{37.46} & \bst{16.84} & \bst{7.59} & \bst{6.93} & 6.92 & 15.8\% \\
~ + joiner gating & 77.2 & 51.78 & 17.57 & \bst{9.42} & 7.45 & 7.06 & \bst{12.02} & \bst{8.43} & \bst{7.41} & 6.94 & 6.83 & \bst{33.85} & \bst{15.13} & \bst{7.43} & \bst{6.83} & 6.83 & 18.7\% \\
~~~~ ++ VIC-Regularization$^\dagger$ & 78.9 & 52.26 & \bst{17.38} & \bst{9.49} & 7.43 & 7.04 & \bst{11.46} & \bst{8.40} & \bst{7.43} & 6.97 & 6.86 & \bst{29.15} & \bst{13.30} & \bst{7.26} & \bst{6.88} & 6.88 & \textbf{19.6\%} \\
\bottomrule
\end{tabular}
}
\vspace{-1.5em}
\end{table*}

\vspace{-0.8em}
\subsection{Implementation details}
\vspace{-0.5em}

All implementations used an in-house extension of the PyTorch-based~\cite{Paszke2019PyTorchAI} \textit{fairseq} toolkit.
We used 80-dimensional log Mel filterbank features that are first projected to 128, then spliced and stacked to 512 dimensions, reducing the sequence length by 4x.
The encoder consists of Emformer blocks with 8 attention heads and a 2048-dimensional feed-forward layer.
The prediction network contains three 512-dimensional LSTM layers with layer-norm and dropout.
Both the encoder and predictor outputs are projected to 1024 dimensions before passing to an additive joiner, which contains a linear layer with $\vert\bar{\mathcal{Y}}\vert = 4097$ output BPE units.
We fixed context embedding dimensionality $D = 256$ for all our experiments.
For joiner gating, we used 4-frame subsegments with left and right contexts of 32 and 4 frames, respectively.
%
%
We trained with an alignment-restricted transducer loss where the alignments were obtained using an HMM-based aligner~\cite{Mahadeokar2021AlignmentRS}.
The auxiliary loss weights were set as: $\lambda_{\mathrm{FR}}=0.1$, $\gamma=1.0$, $\mu=1.0$, and $\nu=0.05$ (see \S~\ref{sec:aux} for a discussion). For feature reconstruction, the chenone labels were obtained using a bootstrapped hybrid ASR system~\cite{Le2019FromST}.
The output dimensionality of the expander network in VIC regularization was set to $\mathfrak{D}=1024$.
All hyperparameters were tuned on the \texttt{dev} set. 
The models were trained on 32 GPUs for 120 epochs with a warm-up of 10k steps, using the Adam optimizer~\cite{Kingma2015AdamAM}.

\vspace{-1em}
\section{Results \& Discussion}

\vspace{-0.5em}
\subsection{Performance of baseline models}
\vspace{-0.5em}

%
Table~\ref{tab:results} compares our proposed method against the baselines on the synthetic \texttt{test} set. 
The Emformer models trained with background augmentation achieved reasonable WERs for high SNR conditions, but failed miserably on low SNR.
For example, Emformer-base ($\boldsymbol{\Lambda}$) obtained 65.7\% on the 100\% shift with SNR of 1 dB. 
Insertion errors caused by trailing background were the biggest contributors in this case, e.g., contributing 77.5\% of the overall WER.

AMS improved WERs for low SNR conditions; for example, 65.7\% $\rightarrow$ 11.2\% on 1 dB, 100\% shift. However, it led to significant regressions for 0\% shift and high SNR cases (shown in \wst{red}), indicating that it is not robust to noisy anchor segments. This is undesirable since such conditions may constitute a major fraction of use cases in device-directed speech. Using concatenation (AMC) instead of subtraction alleviated this issue, improving the relative WER reduction (WERR) from 10.1\% to 14.5\%, on average.

\vspace{-0.9em}
\subsection{Effect of biasing methods}
\vspace{-0.6em}

\begin{figure}[t]
\begin{subfigure}{0.49\linewidth}
\centering
\includegraphics[width=\linewidth]{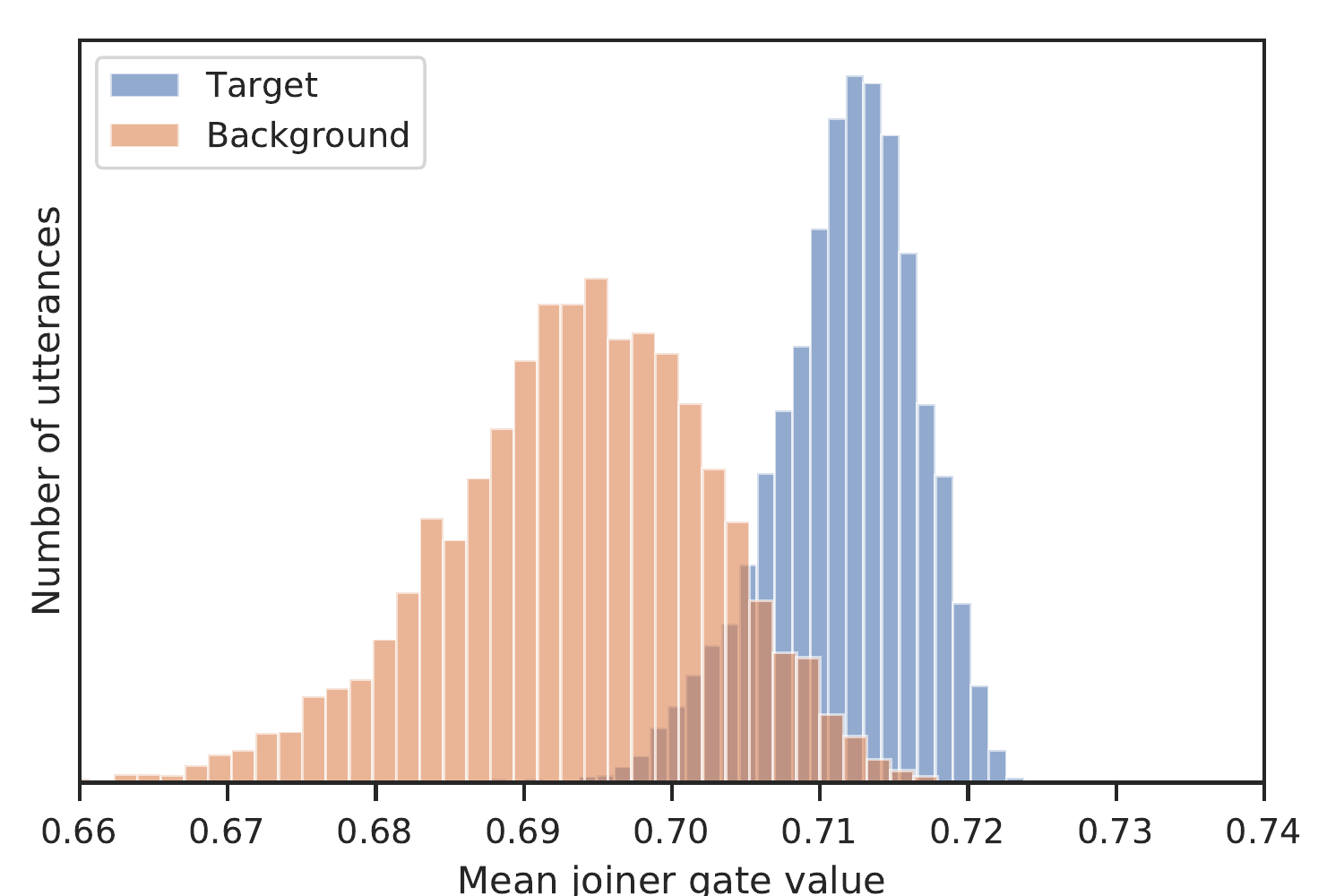}
\end{subfigure}
\begin{subfigure}{0.49\linewidth}
\centering
\includegraphics[width=\linewidth]{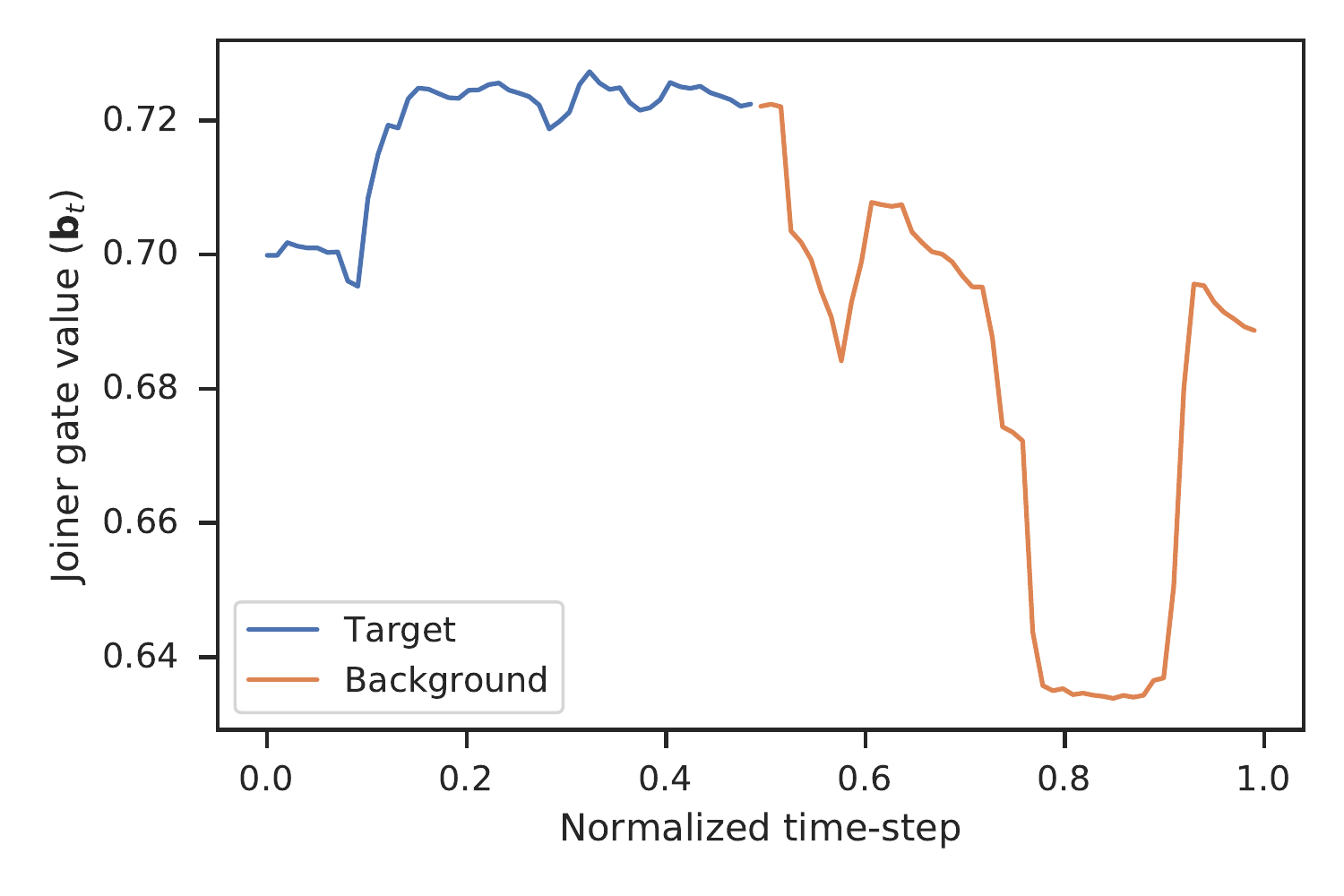}
\end{subfigure}
\vspace{-0.8em}
\caption{
Joiner gate ($\mathbf{b}_t$) values for \texttt{dev} 1 dB, 100\% shift condition. Average over all utterances (\textit{left}) shows clear separation between the modes for target and background segments, which can also be seen in one example utterance (\textit{right}).
}
\label{fig:joiner_gating}
\vspace{-1.5em}
\end{figure}

Both encoder biasing and joiner gating provided WER improvements over $\boldsymbol{\Lambda}$, resulting in average WERR of 15.8\% and 18.7\%, respectively.
Unlike AMS, they did not degrade performance on 0\% shift condition, indicating that they are robust to noisy anchor segments.
Joiner gating helped most on low SNR, high shift conditions by reducing insertion errors.
To analyze the effect of joiner gating, we computed the per-frame similarity scores, $\mathbf{b}_t$ for the \texttt{dev} set for 1 dB SNR and 100\% shift condition, as shown in Fig.~\ref{fig:joiner_gating}.
Since the shift is 100\%, the first half of all utterances contains target speech, and the second half contains background speech (see Fig.~\ref{fig:data_mixing}).
As such, we should observe higher values of $\mathbf{b}_t$ for the first half (shown in \textcolor{blue}{blue}) than the second half (shown in \textcolor{orange}{orange}).
As expected, we see a clear separation between the distribution of similarity scores assigned to the target speech vs. the background speech, which validates our conjecture.
Training with VIC regularization further added a degree of robustness to the model --- the resulting system obtained average WERR of 19.6\%, outperforming all other systems.
Averaged across all SNRs, this model provides WERR of 19.3\% and 36.1\% on the 50\% and 100\% shift cases, respectively, compared to $\boldsymbol{\Lambda}$.

\vspace{-0.8em}
\subsection{Effect of auxiliary objectives}
\vspace{-0.5em}
\label{sec:aux}

\begin{table}[t]
\caption{Effect of auxiliary training objectives on \texttt{dev} set. $\Delta$WER for SNR $>$10 dB was insignificant, and therefore is not shown here. All models use the 20-layer Emformer configuration.}
\label{tab:aux_loss}
\vspace{-0.5em}
\adjustbox{max width=\columnwidth}{%
\begin{tabular}{@{}cc@{\hskip 20pt}ccc@{\hskip 20pt}ccc@{\hskip 20pt}ccc@{}}
\toprule
\multirow{2}{*}{\textbf{FR}} & \multirow{2}{*}{\textbf{VIC}} & \multicolumn{3}{c@{\hskip 20pt}}{\textbf{Shift = 0\%}} & \multicolumn{3}{c@{\hskip 20pt}}{\textbf{Shift = 50\%}} & \multicolumn{3}{c}{\textbf{Shift = 100\%}} \\
\cmidrule(r{20pt}){3-5} \cmidrule(r{20pt}){6-8} \cmidrule(l{-1pt}){9-11}
& & \textbf{1} & \textbf{5} & \textbf{10} & \textbf{1} & \textbf{5} & \textbf{10}  & \textbf{1} & \textbf{5} & \textbf{10} \\
\midrule
\xmark & \xmark & 54.3 & 18.0 & 9.4 & 13.0 & 9.0 & 7.6 & 34.4 & 17.1 & 8.0 \\ 
\cmark & \xmark & 54.6 & 18.1 & 9.5 & 12.9 & \textbf{8.7} & 7.5 & 33.4 & 16.0 & 7.8 \\
\xmark & \cmark & \textbf{54.0} & \textbf{17.7} & \textbf{9.2} & \textbf{12.2} & \textbf{8.7} & 7.5 & \textbf{29.7} & \textbf{14.8} & \textbf{7.6} \\
\cmark & \cmark & 54.7 & 18.9 & 9.6 & 12.3 & \textbf{8.7} & \textbf{7.4} & 30.8 & 16.0 & 7.8 \\
\bottomrule
\end{tabular}
}
\vspace{-1.5em}
\end{table}
Table~\ref{tab:aux_loss} shows ablation studies investigating the impact of auxiliary obectives, as evaluated on the \texttt{dev} set.
We see that although VIC regularization is an unsupervised technique, it outperforms feature reconstruction, particularly on very low SNR cases (1 dB).
This may be because it encourages the context embedding to be more robust to changes in lexical content, pauses, etc. since we force the two halves of the anchor to have similar representations regardless of their chenone labels. 
Regarding VIC coefficients, we found that (i) $\nu$ was largely inconsequential; and (ii) high values of $\gamma$ and $\mu$ improved WERs for high shift cases, but degraded 0\% shift (noisy anchor), indicating that the model was learning to rely too much on the context embedding.
Despite their effectiveness, combining the two objectives, FR and VIC, did not provide any additional gains.
As a result, we used the model trained with VIC regularization to evaluate the \texttt{test} set (shown in Table~\ref{tab:results}).




\vspace{-1em}
\section{Conclusion}
\vspace{-0.5em}

We investigated anchored speech recognition for improving neural transducers in the presence of background speech. A context embedding extracted from the anchor segment was used to bias the model using two complementary biasing methods at the encoder and joiner modules. We also proposed auxiliary training objectives to disentangle style from content in the context embedding. Our methods significantly improved the performance over a strong Emformer-transducer trained with background augmentation, providing average WER reduction of 19.6\% on LibriSpeech mixtures, while being robust to noisy anchor segments.

\textbf{Acknowledgments.} We thank Andros Tjandra, Yuan Shangguan, Ke Li, Leda Sari, and Kaustubh Kalgaonkar for insights and helpful discussions.

\small
\bibliographystyle{IEEEbib}
\bibliography{refs}

\begin{thebibliography}{10}

\bibitem{Graves2012SequenceTW}
Alex Graves,
\newblock ``Sequence transduction with recurrent neural networks,''
\newblock {\em ArXiv}, 2012.

\bibitem{He2019StreamingES}
Yanzhang He, Tara~N. Sainath, Rohit Prabhavalkar, Ian McGraw, Raziel
  {\'A}lvarez, Ding Zhao, David Rybach, Anjuli Kannan, Yonghui Wu, Ruoming
  Pang, Qiao Liang, Deepti Bhatia, Yuan Shangguan, Bo~Li, Golan Pundak,
  Khe~Chai Sim, Tom Bagby, Shuo yiin Chang, Kanishka Rao, and Alexander
  Gruenstein,
\newblock ``Streaming end-to-end speech recognition for mobile devices,''
\newblock in {\em IEEE ICASSP}, 2019.

\bibitem{Wu2020StreamingTA}
Chunyang Wu, Yongqiang Wang, Yangyang Shi, Ching-Feng Yeh, and Frank Zhang,
\newblock ``{Streaming Transformer-Based Acoustic Models Using Self-Attention
  with Augmented Memory},''
\newblock in {\em InterSpeech}, 2020.

\bibitem{Li2020TowardsFA}
Bo~Li, Shuo yiin Chang, Tara~N. Sainath, Ruoming Pang, Yanzhang He, Trevor
  Strohman, and Yonghui Wu,
\newblock ``Towards fast and accurate streaming end-to-end {ASR},''
\newblock in {\em IEEE ICASSP}, 2020.

\bibitem{Shangguan2019OptimizingSR}
Yuan Shangguan, Jian Li, Qiao Liang, Raziel {\'A}lvarez, and Ian McGraw,
\newblock ``Optimizing speech recognition for the edge,''
\newblock in {\em MLSys}, 2019.

\bibitem{Graves2006ConnectionistTC}
Alex Graves, Santiago Fern{\'a}ndez, Faustino~J. Gomez, and J{\"u}rgen
  Schmidhuber,
\newblock ``Connectionist temporal classification: labelling unsegmented
  sequence data with recurrent neural networks,''
\newblock in {\em ICML}, 2006.

\bibitem{Zhang2021BenchmarkingLC}
Xiaohui Zhang, Frank Zhang, Chunxi Liu, Kjell Schubert, Julian Chan, Pradyot
  Prakash, Jun Liu, Ching feng Yeh, Fuchun Peng, Yatharth Saraf, and Geoffrey
  Zweig,
\newblock ``Benchmarking {LF-MMI}, {CTC} and {RNN-T} criteria for streaming
  asr,''
\newblock in {\em IEEE SLT}, 2021.

\bibitem{Li2019ImprovingRT}
Jinyu Li, Rui Zhao, Hu~Hu, and Yifan Gong,
\newblock ``Improving rnn transducer modeling for end-to-end speech
  recognition,''
\newblock in {\em IEEE ASRU}, 2019.

\bibitem{Chan2016ListenAA}
William Chan, Navdeep Jaitly, Quoc~V. Le, and Oriol Vinyals,
\newblock ``Listen, attend and spell: A neural network for large vocabulary
  conversational speech recognition,''
\newblock in {\em IEEE ICASSP}, 2016.

\bibitem{Kim2017JointCB}
Suyoun Kim, Takaaki Hori, and Shinji Watanabe,
\newblock ``Joint {CTC}-attention based end-to-end speech recognition using
  multi-task learning,''
\newblock in {\em IEEE ICASSP}, 2017.

\bibitem{Liu2021ImprovingRT}
Chunxi Liu, Frank Zhang, Duc Le, Suyoun Kim, Yatharth Saraf, and Geoffrey
  Zweig,
\newblock ``Improving {RNN} transducer based asr with auxiliary tasks,''
\newblock in {\em IEEE SLT}, 2021.

\bibitem{Andrusenko2020TowardsAC}
Andrei Andrusenko, Aleksandr Laptev, and Ivan Medennikov,
\newblock ``Towards a competitive end-to-end speech recognition for {CHiME-6}
  dinner party transcription,''
\newblock in {\em InterSpeech}, 2020.

\bibitem{Shi2021ImprovingRT}
Jiatong Shi, Chunlei Zhang, Chao Weng, Shinji Watanabe, Meng Yu, and Dong Yu,
\newblock ``Improving rnn transducer with target speaker extraction and neural
  uncertainty estimation,''
\newblock in {\em IEEE ICASSP}, 2021.

\bibitem{Schwarz2021ImprovingRA}
Andreas Schwarz, Ilya Sklyar, and Simon Wiesler,
\newblock ``Improving rnn-t asr accuracy using context audio,''
\newblock in {\em InterSpeech}, 2021.

\bibitem{Sato2021MultimodalAF}
Hiroshi Sato, Tsubasa Ochiai, Keisuke Kinoshita, Marc Delcroix, Tomohiro
  Nakatani, and Shoko Araki,
\newblock ``Multimodal attention fusion for target speaker extraction,''
\newblock in {\em IEEE SLT}, 2021.

\bibitem{Moriya2022StreamingTA}
Takafumi Moriya, Hiroshi Sato, Tsubasa Ochiai, Marc Delcroix, and Takahiro
  Shinozaki,
\newblock ``Streaming target-speaker {ASR} with neural transducer,''
\newblock in {\em InterSpeech}, 2022.

\bibitem{Wang2020VoiceFilterLiteST}
Quan Wang, Ignacio Lopez-Moreno, Mert Saglam, Kevin~W. Wilson, Alan Chiao,
  Renjie Liu, Yanzhang He, Wei Li, Jason~W. Pelecanos, Marily Nika, and
  Alexander Gruenstein,
\newblock ``{VoiceFilter-Lite}: Streaming targeted voice separation for
  on-device speech recognition,''
\newblock in {\em InterSpeech}, 2020.

\bibitem{Rikhye2021MultiUserVV}
Rajeev~Vijay Rikhye, Quan Wang, Qiao Liang, Yanzhang He, and Ian McGraw,
\newblock ``Multi-user voicefilter-lite via attentive speaker embedding,''
\newblock in {\em IEEE ASRU}, 2021.

\bibitem{Delcroix2018SingleCT}
Marc Delcroix, Kateřina Žmol{\'i}kov{\'a}, Keisuke Kinoshita, Atsunori Ogawa,
  and Tomohiro Nakatani,
\newblock ``Single channel target speaker extraction and recognition with
  {SpeakerBeam},''
\newblock in {\em IEEE ICASSP}, 2018.

\bibitem{molkov2021AuxiliaryLF}
Kateřina Žmol{\'i}kov{\'a}, Marc Delcroix, Desh Raj, Shinji Watanabe, and
  Jan~Honza Cernocky,
\newblock ``Auxiliary loss function for target speech extraction and
  recognition with weak supervision based on speaker characteristics,''
\newblock in {\em InterSpeech}, 2021.

\bibitem{King2017RobustSR}
Brian King, I-Fan Chen, Yonatan Vaizman, Yuzong Liu, Roland Maas, Sree
  Hari~Krishnan Parthasarathi, and Bj{\"o}rn Hoffmeister,
\newblock ``Robust speech recognition via anchor word representations,''
\newblock in {\em InterSpeech}, 2017.

\bibitem{Wang2019EndtoendAS}
Yiming Wang, Xing Fan, I-Fan Chen, Yuzong Liu, Tongfei Chen, and Bj{\"o}rn
  Hoffmeister,
\newblock ``End-to-end anchored speech recognition,''
\newblock in {\em IEEE ICASSP}, 2019.

\bibitem{Shi2021EmformerEM}
Yangyang Shi, Yongqiang Wang, Chunyang Wu, Ching feng Yeh, Julian Chan, Frank
  Zhang, Duc Le, and Michael~L. Seltzer,
\newblock ``Emformer: Efficient memory transformer based acoustic model for low
  latency streaming speech recognition,''
\newblock in {\em IEEE ICASSP}, 2021.

\bibitem{Kim2017DynamicLN}
Taesup Kim, Inchul Song, and Yoshua Bengio,
\newblock ``Dynamic layer normalization for adaptive neural acoustic modeling
  in speech recognition,''
\newblock in {\em InterSpeech}, 2017.

\bibitem{Stafylakis2019SelfsupervisedSE}
Themos Stafylakis, Johan Rohdin, Oldrich Plchot, Petr Mizera, and Luk{\'a}s
  Burget,
\newblock ``Self-supervised speaker embeddings,''
\newblock in {\em InterSpeech}, 2019.

\bibitem{Bardes2021VICRegVR}
Adrien Bardes, Jean Ponce, and Yann LeCun,
\newblock ``{VICReg}: Variance-invariance-covariance regularization for
  self-supervised learning,''
\newblock in {\em ICLR}, 2021.

\bibitem{Panayotov2015LibrispeechAA}
Vassil Panayotov, Guoguo Chen, Daniel Povey, and Sanjeev Khudanpur,
\newblock ``{LibriSpeech}: An {ASR} corpus based on public domain audio
  books,''
\newblock in {\em IEEE ICASSP}, 2015.

\bibitem{Paszke2019PyTorchAI}
Adam Paszke, Sam Gross, Francisco Massa, Adam Lerer, James Bradbury, Gregory
  Chanan, Trevor Killeen, Zeming Lin, Natalia Gimelshein, Luca Antiga, Alban
  Desmaison, Andreas K{\"o}pf, Edward Yang, Zach DeVito, Martin Raison, Alykhan
  Tejani, Sasank Chilamkurthy, Benoit Steiner, Lu~Fang, Junjie Bai, and Soumith
  Chintala,
\newblock ``{PyTorch}: An imperative style, high-performance deep learning
  library,''
\newblock in {\em NeurIPS}, 2019.

\bibitem{Mahadeokar2021AlignmentRS}
Jay Mahadeokar, Yuan Shangguan, Duc Le, Gil Keren, Hang Su, Thong Le, Ching
  feng Yeh, Christian Fuegen, and Michael~L. Seltzer,
\newblock ``Alignment restricted streaming recurrent neural network
  transducer,''
\newblock in {\em IEEE SLT}, 2021.

\bibitem{Le2019FromST}
Duc Le, Xiaohui Zhang, Weiyi Zheng, Christian F{\"u}gen, Geoffrey Zweig, and
  Michael~L. Seltzer,
\newblock ``From senones to chenones: Tied context-dependent graphemes for
  hybrid speech recognition,''
\newblock in {\em IEEE ASRU}, 2019.

\bibitem{Kingma2015AdamAM}
Diederik~P. Kingma and Jimmy Ba,
\newblock ``Adam: A method for stochastic optimization,''
\newblock {\em CoRR}, 2015.

\end{thebibliography}

\end{document}